# New Concept in Moisture detection with Unconventional pore morphology Design

Kusum Sharma, Noor Alam, S. S. Islam
Centre for Nanoscience and Nanotechnology,
Jamia Millia Islamia (A Central University), New Delhi – 110025, INDIA
(*) Corresponding Author Email: sislam@jmi.ac.in
Phone No.: +91 11 26987153

**Abstract:** A break in traditional pore morphology approach, is presented here to see its niche merit over the conventional sensors for water vapour detection. Tubular pores were replaced with normal cone for trace- and inverse cone for RH- level detection. The normal conical pore was fabricated by sheer manipulation of reaction rates of electrolytes, anodic polarization rate and time; and the procedure made reversed in case of inverse cone structure. Sensor with normal cone geometry exhibits response in ppm level with sensitivity of 13pF/ppm, lower detection limit(LOD)~120 ppm with excellent response/recovery time. Lowering LOD further requires alteration of conical geometric parameters in tandem with kinetic theory of water vapour molecules. In contrast, sensor developed from inverse conical structure shows response in RH level and LOD touches down to even less than 20 RH% unlike 45 RH% in conventional RH sensors. Linear response characteristics with sensitivity of 5.14 pF/RH%; surprisingly, the limitations such as nonlinear response, large response recovery time and high hysteresis as observed in conventional anodic alumina based humidity sensors have been removed.

Sensing mechanism in both the structures have been suitably demonstrated and ratified with experimental data. Trace level detection is interpreted with the statistical probabilistic approach in the light of kinetic theory of gases and Brownian energy. A correlation between top surface pore diameter (through which water molecule enters) and the optimized mean free path of vapour molecule is established, and demonstrated its effectiveness for humidity detection in trace level. Results are encouraging and same concept may be tried for detection of other gaseous stimuli including organic vapours.

**Keywords:** conical pore; multi-step anodization; RH; trace; probabilistic mean path.

## 1. Introduction

Unlike humidity, hardly any sensor touches human life so much to meet its increasing demand in day to day life. Highly accurate and reliable moisture detection in different environments becomes more and more important for environmental protection, technological processes, scientific R&D work, natural gas processing, healthcare, pharmaceutical industries, semiconductor-, and food processing industries [1-8]. Trace level detection is notoriously challenging due to adsorption affinity of humidity on metal and other surfaces [7] albeit potential interference from different gas moieties in the measurement process.

Humidity sensors are broadly classified as Relative Humidity (RH) - and trace level [9]. A sensor, specific either to RH or trace level, does not have any clear cut conceptual as well as technological base as of now. Any hygrometric material be it bulk, film, porous- no matter its shape and size, were used blankly for humidity sensor development. The advent of nanoscience and nanotechnology has given a new impetus to the scientific community to develop improved surface area to volume ratio, engineer the pore morphology, surface passivation techniques etc.; although there is visible absence of distinct thumb rule that can spell out the specific condition(s) to make either a RH sensor or a trace level one. As of now, the sensor technology as a whole, evolves and revolves around the as-grown morphology; it may favour RH or the other one [9-11]. RH sensor does not require any stringent conditions on material processing and technology, and this is the reason it is very cheap; even a hygrometric bulk or film can do that job. Surface morphology is never thought of as an important issue in RH unless and until high precision and sensitivity is required.

Conventional sensor technologies rely on an unique principle i.e. adsorption and condensation of water vapour onto solid sensing surface. These are robust, fairly inexpensive, and are extensively used in inert or nonreactive gas matrices [7]. The technology currently being used for measuring trace moisture, mostly includes spectroscopic methods viz. tunable diode laser absorption spectroscopy, Fourier transform infrared spectroscopy and chilled mirror technology [8-9]. However, sensor based technology that perked up successfully in this regime are capacitive metal oxide sensors; and have potential to revolutionize the sensing domain if exploited using new ideas and upcoming trends in technologies.

Porous anodic alumina has been poised in the past for humidity detection due to its superiority in terms of corrosion resistance, thermal and chemical stability, highly hygroscopic, strong mechanical strength, high response and broad range of operation [12]. $Al_2O_3$ based sensors settle excellent scores on some criteri as - (a) wide dynamic range, (b) low price and rugged design, (c) can be calibrated from trace level upto RH, (d) response is unaffected due to any variation in flow rate and temperature upto $100^0C$, (e) can be operated at high pressures, upto approximately 200atm(ca. 20 MPa), and (f) advantage in inline measurements irrespective of variation in temperatures and pressures. Ofcourse, it suffers few drawbacks such as, frequent callibration (every six months) [7] slow response, drift and dormancy issues [13]; hardly anysensor including commercials one that do not suffer more or less such ills. Inspite of that alumina sensors have a faster and more reliable response vis-à-vis others as of today [13-14].

So far, considerable amount of work done on alumina sensors about their performance and sensing mechanism [15-17] formation condition [18-20] and the structure of electrodes [21]. Study on RH sensors mostly dominated the R&D work so far [9-11]. The only available low cost trace moisture sensor reported in recent times has been fabricated by sol-gel route [21-22]. In sol-gel technology, there is no control over the morphology of the porous film except sintering temperature which unlikely to offer a clone even for repeated trials. A possible reason is directly referred to the pores being created by the evaporation of binder, plasticizer and other organic volatile liquid(s) at high temperature leading to size distribution of pores from micro- to nanometer. Sensors developed from such film can even work in RH and ppm level as well.

Serious issues like development of cracks and electrode peel off are there as a common phenomenon [22-25]. Few researchers reported on conical pore structure synthesis for its exclusive application in anti-reflection coating purpose [26].

Trace level sensing therefore, is only possible either with the expensive technology as mentioned and also summarized in Table 1(Supporting information no.SI-1), or we evolve some other simple and low cost methods; nanotechnology in material synthesis and device processing is one such viable option. Interestingly enormous research findings are there; but no idea were emerged on the specific pore morphology required to extend humidity detection either to extremely low RH or even trace level. As per available reports [21-22], trace level sensing mostly banks on cylindrical pore morphology irrespective to pore size and pore size distribution; even microporous structure were found sensing RH [4]. Needless to say, as grown morphologies were not even grown with pre-conceived plan to achieve either RH or the trace level.

In this report, an effort is made to engineer the pore shape geometry in such a way so that selective molecular size will get access inside the pores either for RH- or for trace level measurement separately. This is accomplished by customizing the pore shape into normal cone (for trace level) and inverted cone (for RH); sensing mechanism in both the cases schematically demonstrated and explained in section 4. Two parameters, the backbone behind this concept, are- pore size at the entry point of the moisture and the probable mean free path of the water vapour molecules; play the decisive role in sensing zone selection, whether trace- or RH level. A model based on statistical probabilistic approach comprising kinetic theory of gases and Brownian motion, is used to correlate these parameters. Random Brownian movement, probable Mean free path (MFP) and inelastic scattering loss of energy of the vapour molecules within the pore control the molecular dynamics. It has been found that normal cone whose top pore diameter is of the order of the molecular cluster size, is very sensitive towards trace level detection. As if the pore diameter is acting like a low pass filter selective to small cluster; once it is inside the pore volume, it enjoys enormous freedom unlike tubular structure and condenses on the pore wall surface when Brownian energy reduced to a minimum. Response- and recovery time depends on many factors- inner surface affinity, randomness in its motion, frequency of collisions, and mean free path etc. This is to a great extent not possible in normal tubular shape in a customized fashion as done in the present case; and the reason is explained in section 4.This is the first research article that reports on maneuvering of the pore shape design in porous anodic alumina, and its effectiveness for selective detection either in trace level or RH. Even sensing in the transition region i.e. from trace level to extremely low RH, is also possible and demonstrated as well; and found dependent on the mutual relation among the parameters such as top diameter, molecular cluster size, probabilistic mean free path, and loss of Brownian energy.

The paper is presented in the following sequences: it begins with the description of fabrication technique used to engineer the pore geometry using careful selection of anodization parameters. Secondly, the grown novel nanostructures have been characterized, analyzed and finally exploited for moisture sensing application. The plausible sensing mechanisms at trace- and RH level are explained with the help of statistical probabilistic approach comprising kinetic theory of gases, probable mean free path and Brownian energy. The performance viz. sensitivity, response-and recovery time, hysteresis and cyclic repeatability of the developed sensors for both the structures has been discussed in detail.

## 2. Experimental

A detail methodology is given in **SI**. **Figure 1** schematically describes the formation of segmented porous anodic alumina structure with (a) conical-, and (b) inverse conical pore geometry.

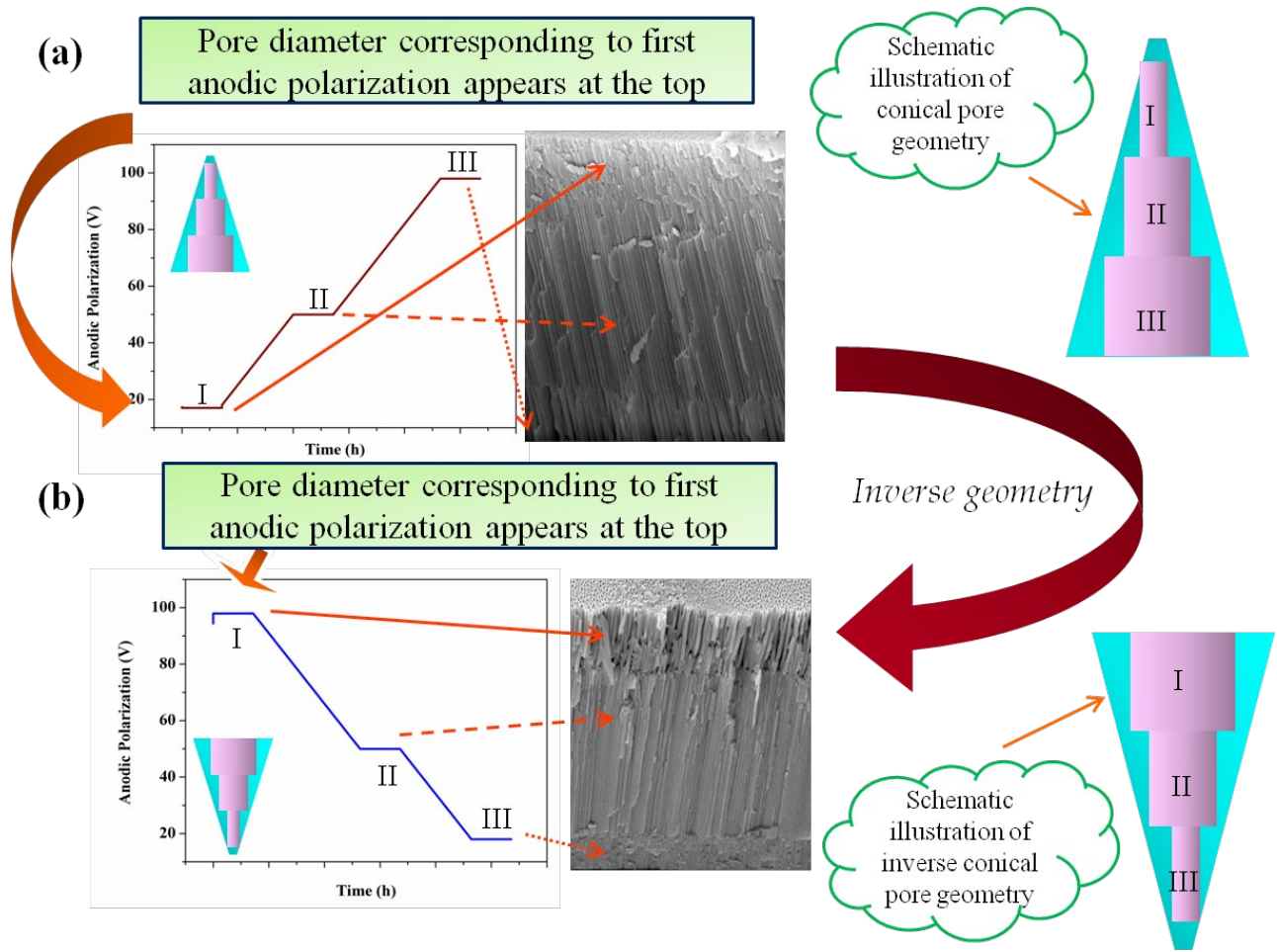

**Figure 1.** Schematic illustration for the fabrication of segmented porous anodic alumina with uncommon geometry - (a) conical, and (b) inverse conical pore geometry.

## 3. Results and discussion

### 3.1. Surface morphology

**Figure 2** shows FESEM image of sample 1(normal conical pore geometry) and 2(inverse conical pore geometry). Pore morphology parameters in both cases are given in Table 3.

**Table 3. Pore morphology parameters in case of (a) sample 1(normal cone), and (b) sample 2 (inverted cone)**

| Geometrical Parameters | | Sample 1(normal cone) | Sample 2 (inverse cone) |
|---|---|---|---|
| **Pore diameter (nm)** | Top | 18 | 140 |
| | Middle | 40 | 37 |
| | Bottom | 90 | 5 |
| **Inter pore spacing (nm)** | Top | 15 | 154 |
| | Middle | 64 | 85 |
| | Bottom | 138 | 30 |
| **Channel Length (μm)** | Top | 0.86 | 2.1 |
| | Middle | 5 | 5.6 |
| | Bottom | 1.4 | 1.2 |
| **Total channel Length (μm)** | | 7.6 | 9 |
| **Barrier layer Thickness (nm)** | | 86 | 5 |

The pore morphology varies directly in accordance with the anoidc polarization. For sample 1 the smallest anodic polarization is applied first and thus the smallest pore diameter is obtained at the top most section. The pore growth starts at the pore bottom while the pore wall was being created; the wall increases in height with time and the barrier layer continually regenrates in the process. With the application of new volatge, the new pore growth starts to occur and the previously formed stucture subsequently shifts upward [18-19].

The pore diameter in sample 1 and Sample 2 changes in three steps as we move from bottom to the top of the pore. Appearance of two segments across the channel length of sample is shown in **Figure 3(a)** for sample 1, and (b) for sample 2 **Figure 3(a'))**.

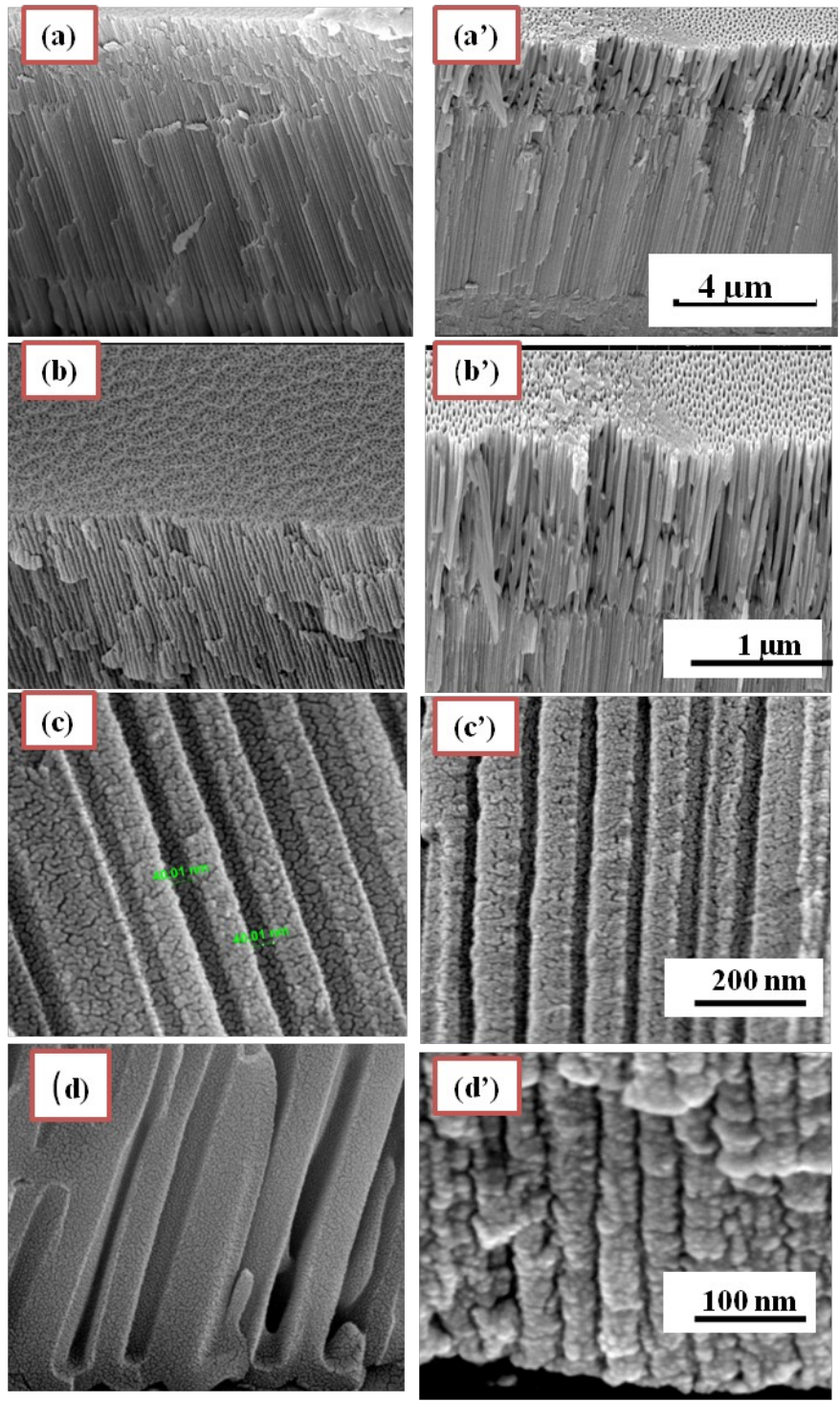

**Figure 2. (a)** FESEM image of complete pore structure in sample 1 (normal cone) where (b) top-section, (c) middle section, (d) bottom section. Similarly, (a') complete pore image of sample 2 (inverse cone), (b') top-section, (c') middle section, (d') bottom section. The scale bar is same for (x-x').

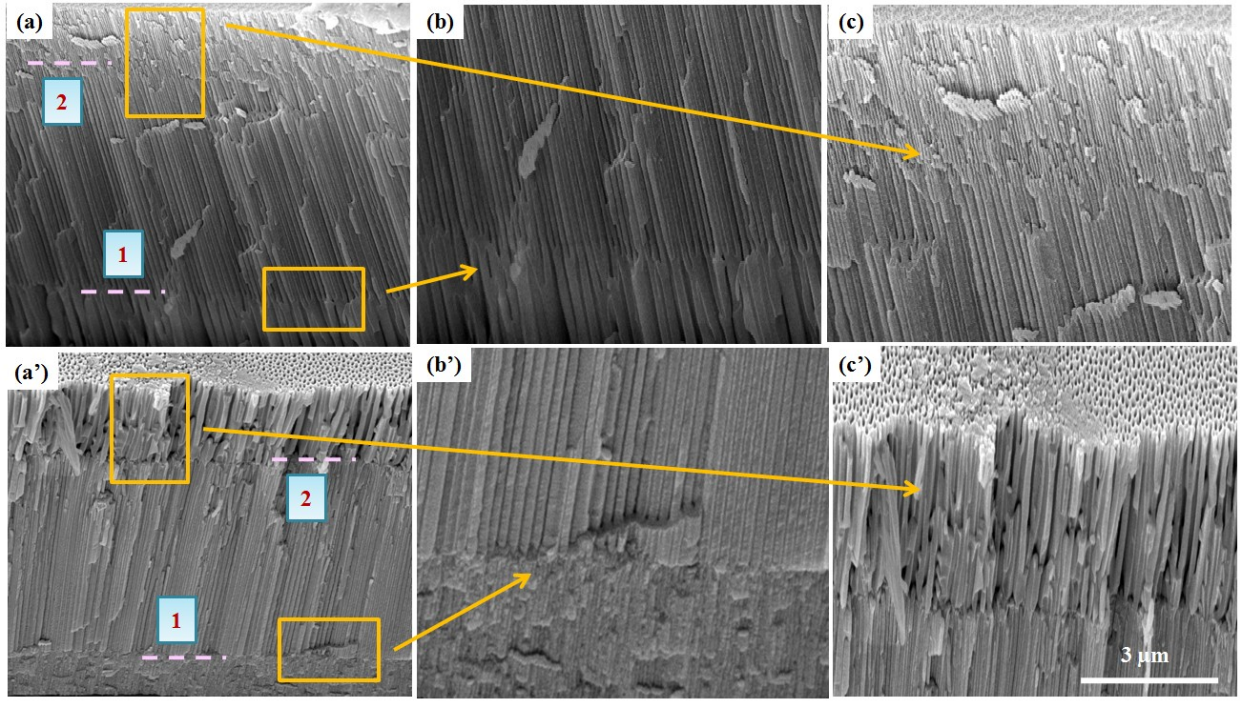

**Figure 3.** (a) The FESEM image of sample 1 depicting step descrease in pore diameter as we move from bottom to the top section, (b) magnified view of segment 1 illustrating the step transition in pore diameter, (c) magnified view of segment 2 illustrating the second step transition in diameter of pore, (a') The FESEM image of sample 2 (inverse pore geometry) describing the step wise increase in pore diameter as we move from bottom to the top section, (b') magnified view of segment 1 illustrating step transition in pore diameter, (c') magnified view of segment 2 illustrating the second step transition in pore diameter of sample 2 (inverse pore morphology design). The scale bar is same for all image.

### 3.2. Electrical characterization of moisture sensor developed from (a) sample 1, and (b) sample 2

Two capacitive sensors were developed where water permeable thin gold layer (2nm) were deposited at the top surface. The humidity sensing measurements were carried out by exposing sensor surface to different humidity concentrations ranging from ppm to RH, where dry commercial grade nitrogen was used as carrier gas, not synthetic air. This is done intentionally as a trial case to establish our concept from sensing performances conducted in the controlled environment of a single carrier gas molecule, and to prevent the interference of other component gases in it. Capacitive sensor response were measured by commercial humidity sensor (Vaisala, HMT 330) and data acquisition by semiconducr characterization system (SCS 4200, Keithley) at 1 KHz. In oder to achieve a stable base capacitance, the sample was purged with dry nitrogen for 12 hours. The sensor was then, exposed to different moisture concentration.

### 3.2 (a) Sensor 1

**Figure 4** shows the capacitive response as a function of humidity. As evident, the sensor with such morphology, is able to detect moisture at trace level, and the lower detection limit is now extended to 120 ppm. The obtained sensitivity was 13 pF/ppm. The capacitive response is linear which is an added advatages for designing the associated circuitary. Such unusual characteristics results from the novel nanostructure fabricated using multi step anodization. Change in dielectric upon adsorption of moisture makes change in capacitive response [27].The sensitivtity is defined in terms of the relative change in capacitance to the corresponding change in moisture concentration [16].

$$S = \frac{C_{max.} - C_{min}}{ppm_{max} - ppm_{min}} X\,100\ pF/ppm \qquad (1)$$

$C_{max.}$ and $C_{min}$ are the maximum and minimum capacitance at respective moisture concentration. The sensor is also checked for high moisture concentration upto 50RH% below which normal RH sensor fails to detect with considerable response. The response characteristics were found to be linear with sensitivity of 33.15pF/RH%, much higher than the reported values [13-15]. It makes the sensor extremely sensititve and can be utilizedeven in application where detection of moisture below 40RH% is critical.

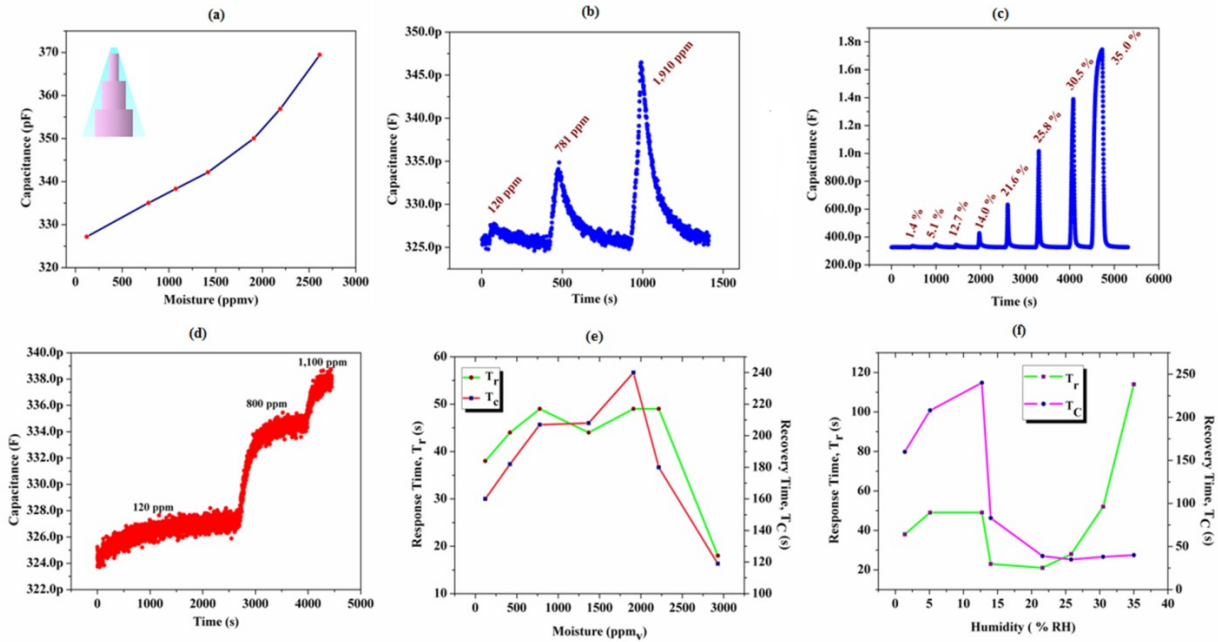

**Figure 4.** (a) Capacitive response of sensor 1 towards moisture concentrations in the range 120 ppm – 2600 ppm and, (b) Static response at low moisture concentration, (c) static response for high moisture concentrations, and (d) dynamic response at low moisture concentration. Response/Recovery plot in the range (e) 120 – 1910 ppm$_v$, and (f) 1.4% - 35% RH%.

### 3.2 (b) *Sensor-2*

**Figure 5(a)** shows the capacitive response as a function of humidity, whereas **Figure 5(b)** shows the capacitive response of a typical porous anodic alumina with cylinderical pores taken as reference for comparative study. Figure demonstrates that the response has been immensely improved in terms of linearity and the lower detection limit extended to even less than 20 RH%, unlike ~ 45 RH% for conventional anodic alumina sensor anodized in oxalic acid.

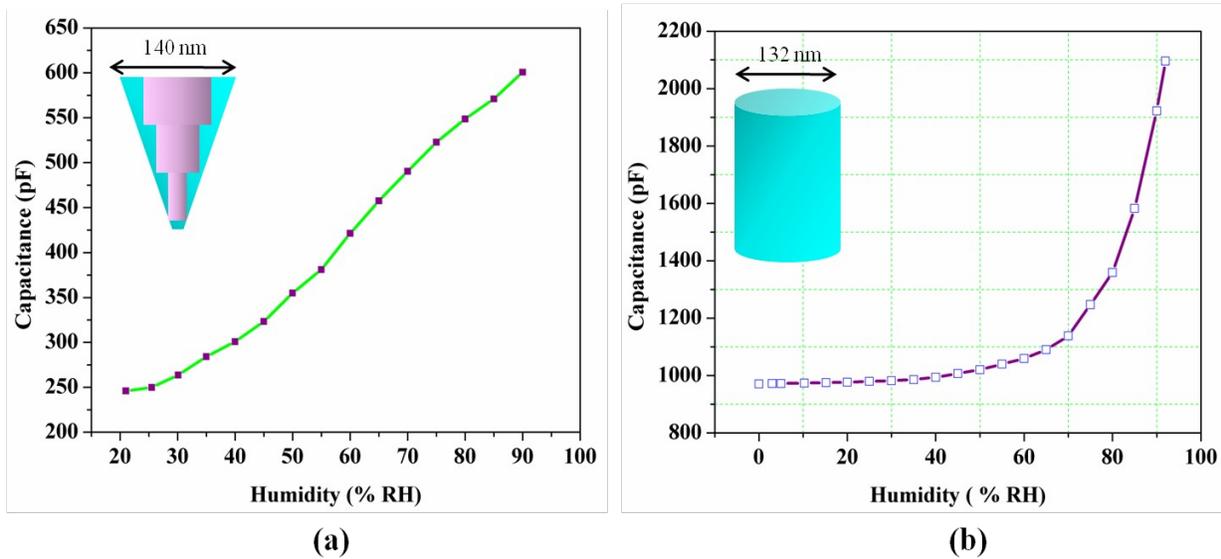

**Figure 5.** (a) Capacitive response of sensor 2 with inverse conical pore geometry where top surface average pore diameter ($d_{av}$) – 140 nm ,(b) Capacitive response of a sample with uniform tubular pore of average diameter ($d_{av}$) = 132 nm.

**Figure 6(a)** shows the static plot of the prepared sensor exhibiting how sentivite the sensor when minor change in concentration is executed. The sensitivity of sensor 2 is 5.14 pF/RH%and is fairly large compared to the reported conventional sensors [9]. The sensor responds extremely fast to the change in moisture at low RH with response- and recovery time 63s and 22s respectively **(Figure 6(b))**. For repeatability check, the sensor was exposed to a step change in moisture 35 times. The sensor shows very small hystersis of ~ 1% in the RH range 20 to 47% **(Figure 6(c))**.

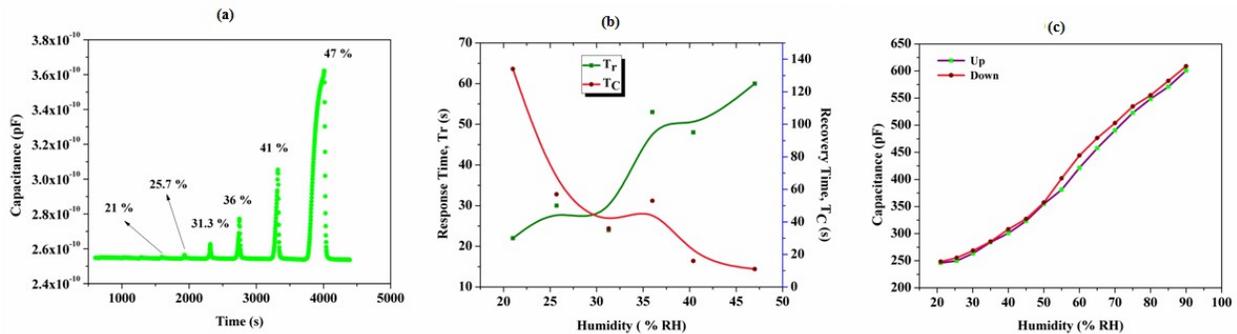

**Figure 6.** (a) static response/recovery plot, (b) static response/recovery plot as a function of humidity and, (c) hysteresis plot of sensor 2 with invesrs conical pore geometry.

## 4. Sensing mechanism

Moisture sensing in porous structure takes place in three consecutive steps –(1) the entry of water molecules through pore, (2) random motion cum relaxation through frequent inelastic collisions leading to reduction in Brownian energy, and finally (3) exit through the entry door; or adsorb on the pore wall surface, subject to the level of Brownian energy of the molecules. On adsorption, the molecules undergo through different chemical and physical processes. Schematic diagram shown in **Figure 7** demonstrate the above facts. An efficient and sensitive sensor does comply these conditions. Adsorption directly relates to surface properties, and condensation happens when water vapour molecules coalesce. Plenty of experimental studies have endorsed the fact that porous alumina ceramic is highly moisture sensitive both in RH and sub-ppm range as well, if its porosity is properly controlled.

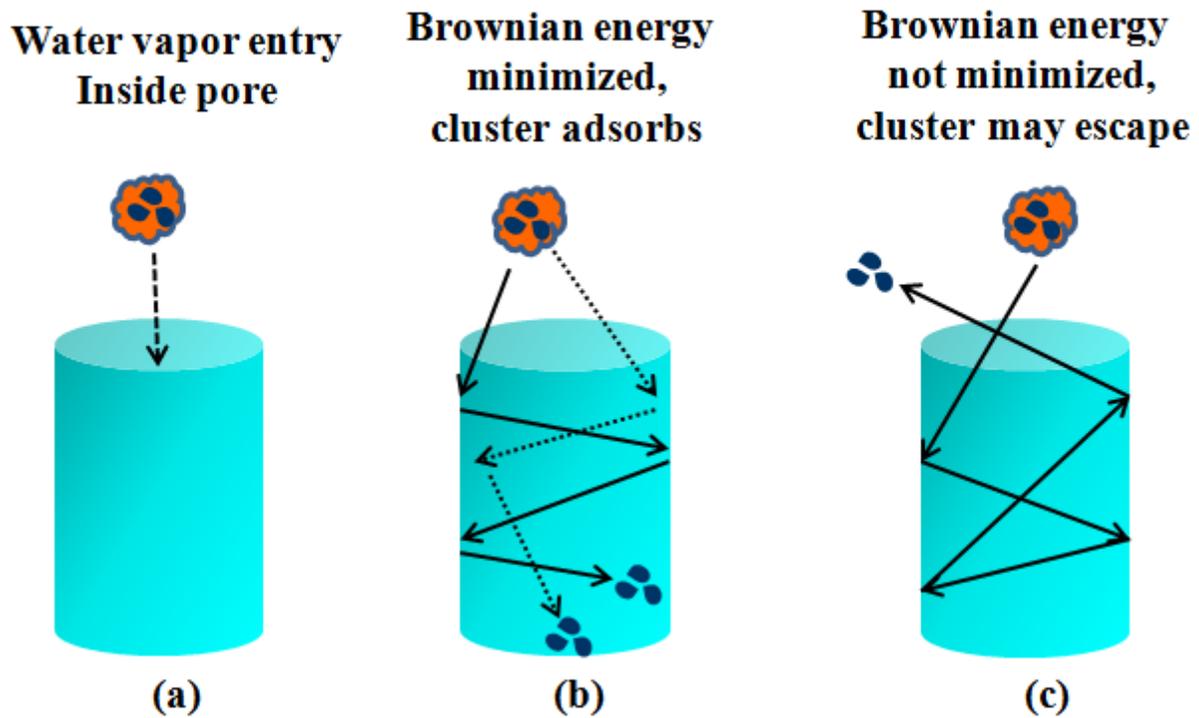

**Figure 7.** Schematic explaining the three consecutive steps of moisture adsorption in conventional (cylindrical) pores.

Pore size distribution and surface area have been widely talked so far, for their decisive role in quantifying sensing parameters irrespective of detection range [28-29]. Three more important aspects, that govern the sensor performance as a whole, remained in oblivion. These are –

(i) *Compatibility of the molecular cluster size with the pore size; a condition essential, but not sufficient. Statistically, equality in size between pore and the cluster, cannot be taken as a thumb rule for cluster entry; large displacement of the cluster near the pore entry may deter the cluster from access inside the pore.*

(ii) *Net molecular displacement at the pore entry; that decides on the molecular entry inside the pore. In other words, the net deviation between the desired pore size and* *mean free path (MFP) of the cluster is to be minimized to allow the cluster to move close to the potential field of the pore entry, and lastly*

(iii) *after entry inside the pore, the next task is how to bring down the Brownian energy of the cluster to a minimum level for its adsorption inside the pore wall surface.*

Therefore, right from pore entry to adsorption, optimization of both, mean free path, $\lambda$ (*MFP*) and pore size *(Q)* is necessary [40]. The compatibility between $\lambda$ and $Q$ is crucial at sub-nano scale cluster size where ppm/ppb/ppt level sensing takes place; $\lambda$ is insignificant for RH because of the wide pore entry i.e. $Q >> \lambda$; it will accommodate all $\lambda$'s. This is discussed at length in section 4B.

*i. Probabilistic Mean free path Optimization:*

Under steady state, the mean free path $\lambda_a$ *(MFP)* of a molecule of radius '$r_a$' (of mass $m_a$) in a gas mixture with another component i.e. carrier gas of molecular radius '$r_b$'(of mass $m_b$) can be written from kinetic theory of gases and Maxwell's equation [30] as

$$\lambda_a = \frac{1}{\pi \left\{ \sqrt{2} n_a (2 r_a)^2 + n_b (r_b + r_a)^2 \sqrt{1 + \frac{m_a}{m_b}} \right\}}$$

(2)

where $\lambda_a$, is the MFP of the water vapor, $n_a$ and $n_b$ are the molecular density of water vapor and carrier gas i.e. nitrogen, $r_a$ and $r_b$ are the respective radii, $m_a$ and $m_b$ are mass of single molecule component of water vapor and nitrogen gas.

It is evident from eq.2 that $\lambda_a$ is an inverse function of molecular size ($r$), mass($m$); partial vapour pressure of individual gas, temperature, and viscosity also affect it [30]. In fact, $\lambda_a$ in eq. 2 treats the molecules as hard spheres, but real molecules are not. Hard sphere approximation works fine in case of noble gases, as the collisions are probably close to being perfectly elastic. But each water vapour molecule is a dipole; develops electrical interaction as they approach each other. This has been taken care by using an electrical potential for the molecules to refine the mean free path calculation. Greater the attractive forces between them, more the gas will deviate from ideal gas behavior. In this perspective, eq. 2 is erroneous for any gas having polar molecule. To overcome this error, we will treat it as probabilistic mean free path '$\lambda_p$' instead of '$\lambda_a$', and '$\lambda_p$'will be equally probable for sensing at all ppm levels subject to the availability of compatible pore size *(Q)* **(Figure 8(c(i)))** [31] . The equality in size between pore and the cluster, cannot be taken as a thumb rule for cluster entry; large molecular displacement of the cluster at the pore mouth may deter the cluster from access inside the pore **(Figure 8(c(ii)))**. Therefore, $\lambda_p$ can have all possible values from $\lambda_{pmin}$ to $\lambda_{pmax}$; and its uniform distribution is also supported by statistics [30].

The probability density function of the probabilistic MFP $\lambda_p$ of water vapour molecule is [31]

$$f(\lambda_p) = \frac{1}{\lambda_{pmax} - \lambda_{pmin}}, where \lambda_{pmin} \leq \lambda_p \leq \lambda_{pmax}; f(\lambda_p) > 0 \qquad (3)$$

for $\lambda_p < Q$, the probable displacement of the molecule near to pore mouth is $(Q-\lambda_p)f(\lambda_p)$ and for $\lambda_p > Q$ it is $(\lambda_p-Q)f(\lambda_p)$. Therefore, the total probable displacement D suffered by the vapour molecule will be the sum total of individual displacement integrated over the $\lambda_p$ range as

$$D(Q) = \int_{\lambda_{pmin}}^{Q} (Q-\lambda_p)f(\lambda_p)d\lambda_p + \int_{Q}^{\lambda_{pmax}} (\lambda_p-Q)f(\lambda_p)d\lambda_p \qquad (4)$$

*ii. Pore size Optimization*

$D(Q)$ will be minimum when the distributions of $\lambda_p$ is to be at the least deviation from the desired pore size $Q$ and it ensures $\lambda_{p\text{-}opt}$ and Q be comparable. The pore size $(Q)$ optimization may be done by taking first derivative of $D(Q)$ and equate to zero [31].

$$\frac{dD(Q)}{dQ} = \int_{\lambda_{pmin}}^{Q} (1-0)f(\lambda_p)d\lambda_p + \left[(Q-\lambda_p)f(\lambda_p)\frac{d\lambda_p}{dQ}\right]_{\lambda_p=\lambda_{pmin}}^{Q}$$

$$+ \int_{Q}^{\lambda_{pmax}} (0-1)f(\lambda_p)d\lambda_p + \left[(\lambda_p-Q)f(\lambda_p)\frac{d\lambda_p}{dQ}\right]_{\lambda_p=Q}^{\lambda_p=\lambda_{pmax}}$$

$$¿ \int_{\lambda_{pmin}}^{Q} f(\lambda_p)d\lambda_p - \int_{Q}^{\lambda p_{max}} f(\lambda_p)d\lambda_p$$

$$¿ \int_{\lambda_{pmin}}^{Q} f(\lambda_p)d\lambda_p - \left[1 - \int_{\lambda_{pmin}}^{Q} f(\lambda_p)d\lambda_p\right]\left\{\forall \int_{\lambda_{pmin}}^{\lambda_{pmax}} f(\lambda_p)d\lambda_p = 1\right\}$$

$$¿ 2 \times \int_{\lambda_{pmin}}^{Q} f(\lambda_p)d\lambda_p - 1 = 0 \Rightarrow \int_{\lambda_{pmin}}^{Q} f(\lambda_p)d\lambda_p = \frac{1}{2} \qquad (5)$$

On substitution of density function $f(\lambda_p)$ from eq. 3, the first term in eq. 4 will be

$$\int_{\lambda_{pmin}}^{Q} \frac{1}{\lambda_{pmax} - \lambda_{pmin}} d\lambda_p = \frac{1}{2} \qquad (6)$$

and optimized pore size (Q) is then obtained from eq.5 as

$$Q_{opt} = \frac{1}{2}(\lambda_{pmax} - \lambda_{pmin}) + \lambda_{pmin} \qquad (7)$$

When $Q$ becomes $Q_{opt}$, net deviation of the vapour molecule from the pore mouth attains a minimum value, and $D(Q)$ becomes minimum; rather be named as minimization function. The condition for $D(Q)$ to be a minimum function is [32]

$$\frac{d^2 D(Q)}{d\lambda_p^2} = 2\left[f(\lambda_p)\left(\frac{d\lambda_p}{dQ}\right)\right]_{\lambda_{pmin}}^Q = 2f(Q) > 0 \tag{8}$$

Conventional pore geometry so far investigated is cylindrical type; it is readily available in sizes, and comparatively easier to model vis-à-vis non-uniform or other complex shapes using Molecular Dynamics (MD) simulations [33-36]. Obviously, it attaches scientific importance to execute MD simulations to study the dynamics of water vapour molecules in complex pore systems, not only for science but applications as well. The present study deals with complex pore geometry i.e. conical and inverse conical shape design as for trial investigation. A comprehensive study shows that both two structures are individually specific to specific range of humidity detection.

### A. *For normal conical pore geometry*

**Figure 8** schematically describes the entry and adsorption of water vapour molecule in conical pore geometry. Water vapor molecules enter through the top surface that has smaller diameter with respective to the bottom surface. The top surface acts like a low pass filter in electronic circuits, and screening of cluster sizes comparable to top pore diameter $(d_u)$ takes place at the first instance **(Figure 8(b))**. At low humidity concentration i.e. trace level, the cluster size as well as $\lambda_P$ both are of the order of nanometer and can get access inside the pore [31]. Once it is in, the following four things may happen within the pore volume –

- At ambient condition, inter molecular collisions and their random movements within confined space is termed as Brownian motion. Since the pore volume is quite large, frequent collision is less probable resulting in large mean free path, $\lambda_p$; less kinetic energy losses due to inelastic scattering. The situation is similar to 'Blackbody radiation' phenomenon – the photon enters through a small hole and wanders throughout the volume, and not able to escape easily **(Figure 8(d))** and because of that the recovery is slow at low moisture concentration.
- Inside the pore, Brownian movement of the clusters will be restricted by- (1) the inelastic scattering with other molecules and the pore walls, and (2) dipole–dipole interaction between the polar molecules. The stronger the second factor, the more the adsorption and condensation on the surface, resulting in slow response- and large recovery time. There is close match between theory and experiment and is shown in **Figure 4(e).**
- At humidity concentration of the order of sub-ppm or even less than that, the cluster may adsorb eventually due to the reason mentioned. More and more similar clusters will join and make a bigger cluster. As a result, detection level may extend to even higher ppm to sub-RH. Experimental results shown in **Figure 4(c)** also support this approach.
- Water molecules first chemisorbs on the available sites by forming hydrogen bonds with the oxygen atoms of the oxide surface. At low humidity, after the first layer of water molecules get adsorbed on the surface, a dissociative mechanism leads to the formation of a hydroxyl ion ($OH^-$) and a proton ($H^+$) **(Figure 8(e))** [37]. With increasing humidity, further physisorbed layers are formed, then $H^+$ ion can move freely in the physisorbed water according to the Grothuss's chain reaction [38-39].

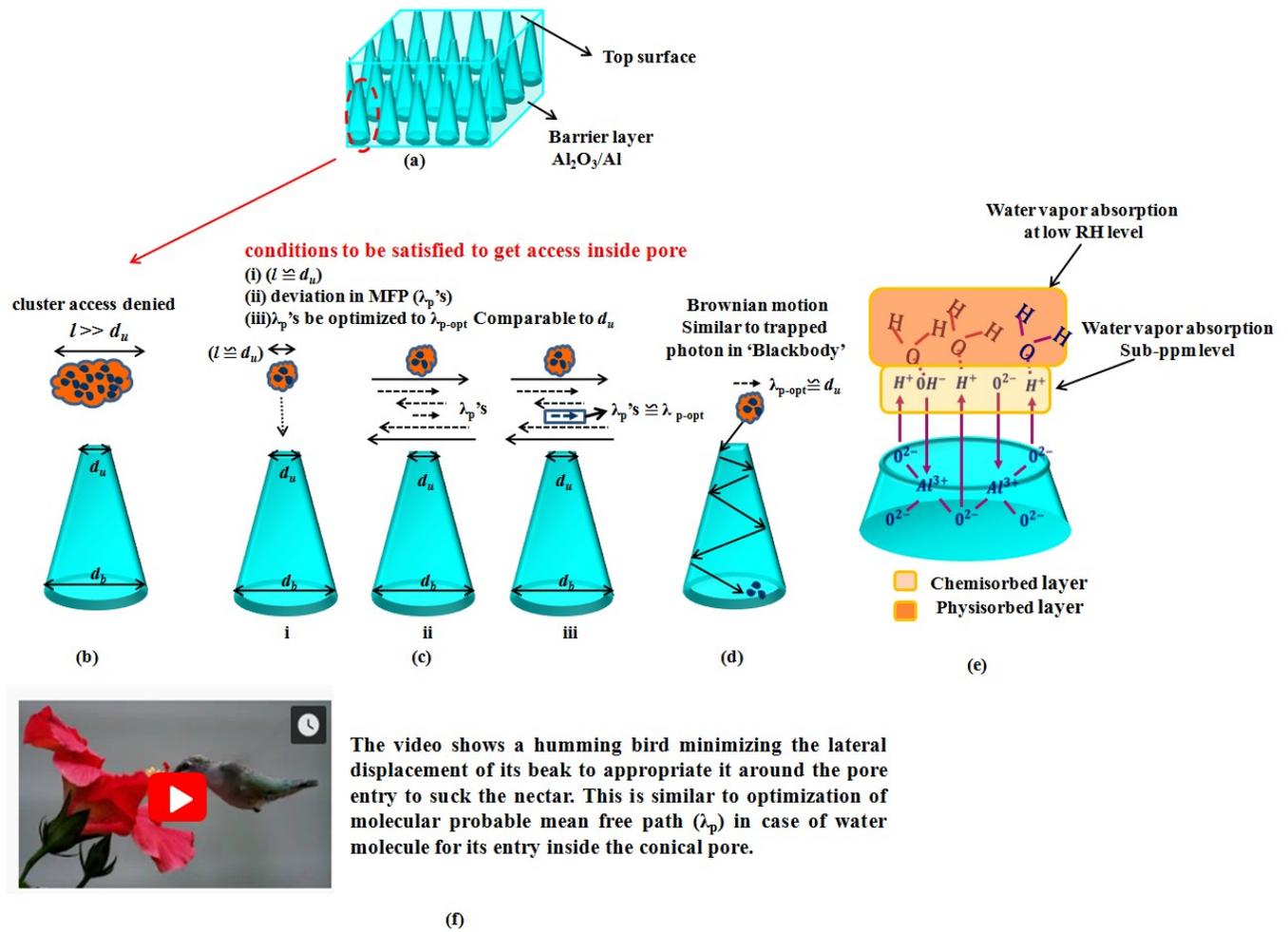

**Figure 8.** Schematic showing (a) 3D-view of porous anodic alumina with conical pore geometry, (b) why RH level sensing is not possible in such morphology, (c) conditions of molecular cluster entry inside pore (i) compatibility between molecule cluster size and pore size, (ii) molecule cluster deviation at pore mouth, (iii) cluster entry inside pore at probabilistic MFP (d) cluster dynamics once inside the pore, (d) water vapor adsorption at sub-ppm and low RH level and, (g) video demo showing bio-immitation of a humming bird sucking nectar from flower inside akin to water molecular cluster entry inside conical pore geometry.
 (https://youtu.be/8MWVDA7gDGw).

**B.** *For Inverse conical pore geometry*

In inverse cone, the top surface through which water molecules enter have larger diameter $(d_u)$ vis-à-vis the bottom surface $(d_b)$ as shown in **Figure 9.** Clusters, no matter what the size is, will get easy access through the top surface. There will be no restriction on the cluster size so long the top surface diameter of the cone is much bigger; even cluster size comparable to high RH will get free entry. Let us consider one smallest cluster of size *l*. Since $l << d_u$, cluster will enter inside the pore, bounce to and fro multiple times from the wide surface wall. The mean free path, $\lambda_p$ in this case will be very high as the cluster finds itself much more free than the tubular uniform pore structure since the cone volume is much bigger than that of tubuler pore of same dimension. Besides, less inelastic scattering with other clusters renders little loss of Brownian energy leading its easy escape through the wide entry root **(Figure 9(b))**. Not only that the cluster has little probability to adsorb on the wall unless and until the cluster becomes bigger as comparable to sub-RH level **(Figure 9(c))**. This is the reason the trace level sensing is not possible in such structure. since the escape out is an easy phenomenon and sensing response is virtually negligible.

With increase in vapour concentration, the cluster size becomes bigger and bulky by weight. At cluster sizes comparable to RH level, frequency of collision do also increase. As a result, scattering losses becomes high, reduction of Brownian energy is more, and cluster escape is not too easy unless an external arrangement is made to boost the Brownian energy of the cluster. In that case, the recovery will be a slow process and RH level sensing will dominate more and more **(Figure 9(d))**. Initially, response- and recovery process will be faster; when RH sensing dominates, response remains nearly unchanged but recovery will be too lengthy. The results are in good agreement with the response- and recovery time as a function of humidity concentration as shown in **Figure 6.**

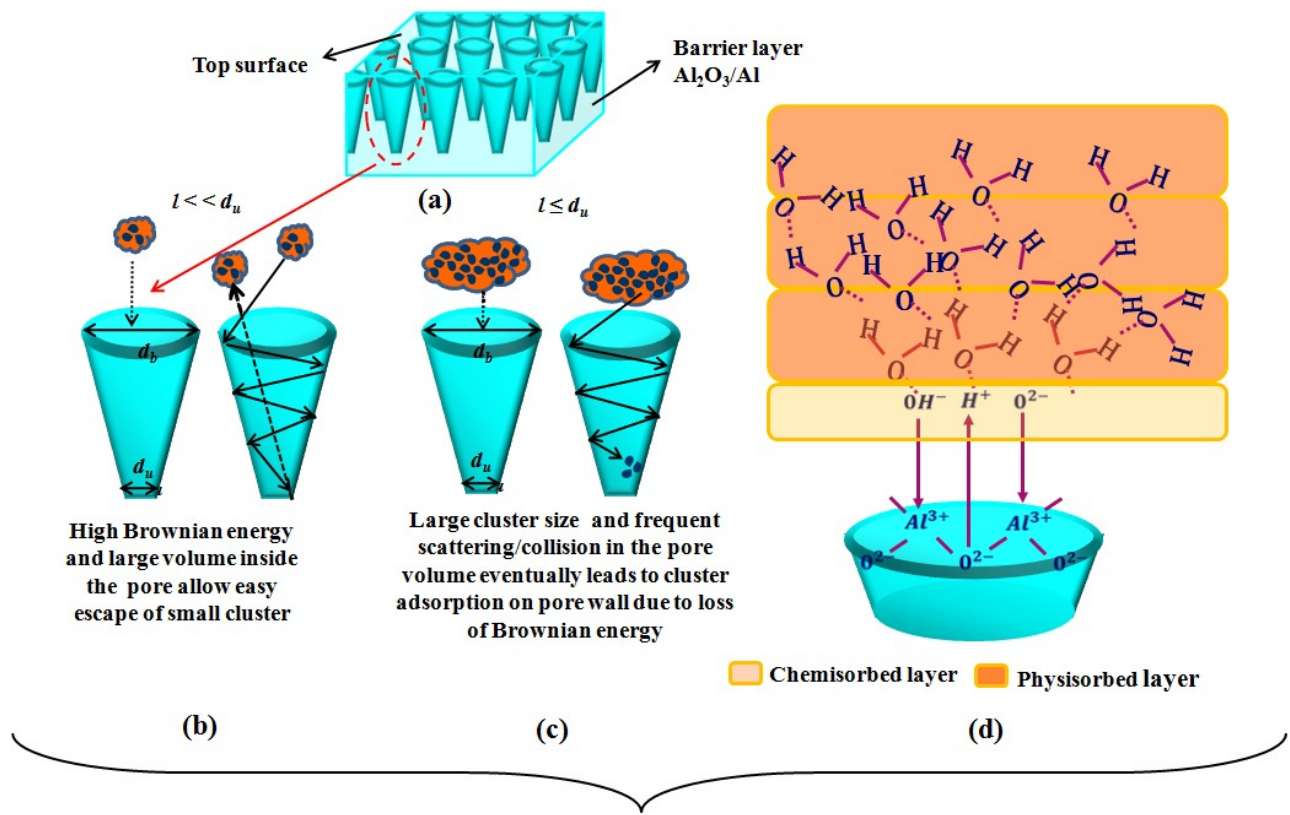

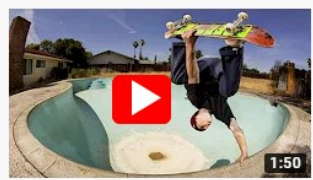

**Figure 9.** Schematic demonstration- (a) 3D-view of porous anodic alumina in inverse conical pore geometry, (b) exit of small clusters due to high Brownian energy, (c) cluster enetry and its dynamics inside the pore at sub-RH level, (d) moiture adsorption at sub RH level and (d) a video demo showing the bio-immitation of molecular cluster with the movement of the skater. (https://www.youtube.com/watch?v=StCh3dlLa8k).

**Conclusion**

Trace- as well as RH level moisture sensors were developed with customized conical pore structure; this is in contrast to the cylindrical pore concept used in mostly available solid state sensors that are unable to detect at trace level. This is accomplished by an unconventional approach where conventional tubular pore structure were replaced with normal cone for trace- and inverse cone for RH level detection. There is a marked difference as well as advantages how the developed sensor overcomes the limitations such as nonlinearity in response, high response- and recovery time, and high hysteresis observed in conventional anodic alumina sensor. In case of normal conical pore geometry, the detection limit touched down to 120 ppm with excellent sensitivity ~13 pF/ppm; further the response characteristics are linear with sensitivity of 5.14 pF/RH% for inverse conical pore geometry. Trace level sensing in normal cone structure has been suitably demonstrated and interpreted with the statistical probabilistic approach in the light of kinetic theory of gases and Brownian motion. A correlation between top surface pore diameter (through which water molecule enters), the molecular cluster size and its optimized probable mean free path is established and showed its effectiveness for humidity sensing in sub ppm level.

Such approach is not rigorous for sensing in inverted cone structure; so long the top surface diameter is much bigger than the bottom one, it will not stop any molecule subject to cluster size fits within pore wall. Small clusters will escape out due its higher Brownian energy, whereas the big clusters comparable to the size of RH level will attach on the wall surface and condense later. Literature survey confirms this report as the first one on anodic alumina sensor with normal- and inverted conical pore design with selective response in trace- as well as RH level in a customized fashion. These studies were carried out multiple times to check any disagreement between the concept used and the experimental results. Some more studies are in theprocess to make the sensor more sensitive to extend LOD in ppb/ ppt level in case of trace level sensor and same for RH one, using geometrical parameters of the cone structure such as diameter ratio of pores between top and bottom, number of anodization steps within a pore, increase/decrease in pore channel length etc. As of now, the trial results are encouraging and the same concept may be used for other gas moieties too.

**Acknowledgment**

This research work is funded by Directorate of Extramural Research & Intellectual Property Rights, Defence Research and Development Organization(DRDO), India (No. ERIP/ER/DG-MED &CoS/990216503/M/01/1620).

**References**

1) Wang, Hongxia, et al. "Multifunctional Directional Water-transport Fabrics with Moisture Sensing Capability." ACS applied materials & interfaces (2019).
2) Kumar, Rajat, Ramesh Naidu Jenjeti, and Srinivasan Sampath. "Bulk and Few-Layer 2D, p-MnPS3 for Sensitive and Selective Moisture Sensing." Advanced Materials Interfaces (2019): 1900666.
3) Karthikeyan, S., et al. "Graphene oxide-based optical waveguide for moisture sensing in transformer oil." Photonics and Nanostructures-Fundamentals and Applications 36 (2019): 100727.
4) Iyengar, Sathvik Ajay, et al. "Surface Treated Nanofibers for High Current Yielding Breath Humidity Sensors for Wearable Electronics." ACS Applied Electronic Materials (2019).
5) Alrammouz, R., et al. "Highly porous and flexible capacitive humidity sensor based on self-assembled graphene oxide sheets on a paper substrate." Sensors and Actuators B: Chemical 298 (2019): 126892.
6) Andika, Rachmat, et al. "Organic nanostructure sensing layer developed by AAO template for the application in humidity sensors." Journal of Materials Science: Materials in Electronics 30.3 (2019): 2382-2388.

7) Raynor, Mark W., et al. "Trace water vapor analysis in specialty gases: sensor and spectroscopic approaches." Trace Analysis of Specialty and Electronic Gases, WM Geiger and MW Raynor, Eds (2013): 195-249.
8) Funke, Hans H., et al. "Techniques for the measurement of trace moisture in high-purity electronic specialty gases." Review of scientific instruments 74.9 (2003): 3909-3933.
9) Chen, Zhi, and Chi Lu. "Humidity sensors: a review of materials and mechanisms." Sensor letters 3.4 (2005): 274-295.
10) Dickey, Elizabeth, et al. "Room temperature ammonia and humidity sensing using highly ordered nanoporous alumina films." Sensors 2.3 (2002): 91-110.
11) Gong, Dawei, et al. "Highly ordered nanoporous alumina films: Effect of pore size and uniformity on sensing performance." Journal of materials research 17.5 (2002): 1162-1171.
12) Gangwar, Jitendra, et al. "Phase dependent thermal and spectroscopic responses of $Al_2O_3$ nanostructures with different morphogenesis." Nanoscale 7.32 (2015): 13313-13344.
13) Nahar, R. K. "Study of the performance degradation of thin film aluminum oxide sensor at high humidity." Sensors and Actuators B: Chemical 63.1-2 (2000): 49-54.
14) Chen, S. W., et al. "Sensitivity evolution and enhancement mechanism of porous anodic aluminum oxide humidity sensor using magnetic field." Sensors and Actuators B: Chemical 199 (2014): 384-388.
15) Juhász, L., and J. Mizsei. "Humidity sensor structures with thin film porous alumina for on-chip integration." Thin Solid Films 517.22 (2009): 6198-6201.
16) Sharma, Kusum, and S. S. Islam. "Optimization of porous anodic alumina nanostructure for ultra high sensitive humidity sensor." Sensors and Actuators B: Chemical 237 (2016): 443-451.
17) Zhao, Huaizhou, et al. "Ultrasensitive chemical detection using a nanocoax sensor." ACS nano 6.4 (2012): 3171-3178.
18) Lee, Woo, and Sang-Joon Park. "Porous anodic aluminum oxide: anodization and templated synthesis of functional nanostructures." Chemical reviews 114.15 (2014): 7487-7556.
19) Zhou, Zimu, and Stephen S. Nonnenmann. "Progress in Nanoporous Templates: Beyond Anodic Aluminum Oxide and Towards Functional Complex Materials." Materials 12.16 (2019): 2535.
20) Leontiev, A. P., I. V. Roslyakov, and K. S. Napolskii. "Complex influence of temperature on oxalic acid anodizing of aluminium." ElectrochimicaActa (2019).
21) Islam, Tarikul, et al. "A micro interdigitated thin film metal oxide capacitive sensor for measuring moisture in the range of 175–625 ppm." Sensors and Actuators B: Chemical 221 (2015): 357-364.
22) Pandey, Manju, et al. "Development of commercial trace moisture sensor: a detailed comparative study on microstructural and impedance measurements of two phases of alumina." Electronic Materials Letters 10.2 (2014): 357-362.

23) Pandey, Manju, et al. "Nanoporous alumina (γ-and α-phase) gel cast thick film for the development of trace moisture sensor." Journal of sol-gel science and technology 68.2 (2013): 317-323.
24) Pandey, Manju, et al. "Polymer optimization for the development of low-cost moisture sensor based on nanoporous alumina thin film." Journal of Advanced Ceramics 2.4 (2013): 341-346.
25) Pandey, Manju, et al. "Nanoporous morphology of alumina films prepared by sol–gel dip coating method on alumina substrate." Journal of sol-gel science and technology 64.2 (2012): 282-288.
26) Huang, Li-Feng, et al. "Graded index profile of anodic alumina films that is induced by conical pores." Applied optics 32.12 (1993): 2039-2044.
27) Pandey, Manju, Kusum Sharma, and Saikh Safiul Islam. "Wide Range RH Detection with Digital Readout: Niche Superiority in Terms of Its Exceptional Performance and Inexpensive Technology." Advances in Materials Physics and Chemistry 9.2 (2019): 11-24.
28) Zhang, Shumin, et al. "Facile fabrication of a well-ordered porous Cu-doped SnO2 thin film for H2S sensing." ACS applied materials & interfaces 6.17 (2014): 14975-14980.
29) Jiang, Kai, et al. "Excellent humidity sensor based on LiCl loaded hierarchically porous polymeric microspheres." ACS applied materials & interfaces 8.38 (2016): 25529-25534
30) M. N. Saha, and B. N. Srivastava. "A Treatise on Heat, 1950." 3rd Ed. The Indian Press, Calcutta, 1950.
31) Banerjee, Goutam, and Kamalendu Sengupta. "Pore size optimisation of humidity sensor—a probabilistic approach." Sensors and Actuators B: Chemical 86.1 (2002): 34-41.
32) Gillett, Billy E. Introduction to operations research: a computer-oriented algorithmic approach. Tata McGraw-Hill Education, 1979.
33) Gelb, Lev D., et al. "Phase separation in confined systems." Reports on Progress in Physics 62.12 (1999): 1573.
34) Liu, L., et al. "Slow dynamics of supercooled water confined in nanoporous silica materials." Journal of Physics: Condensed Matter 16.45 (2004): S5403.
35) Bourg, Ian C., and Carl I. Steefel. "Molecular dynamics simulations of water structure and diffusion in silica nanopores." The Journal of Physical Chemistry C 116.21 (2012): 11556-11564.
36) S. O. "Pore-size dependence and characteristics of water diffusion in slit like micropores." Physical Review E 92.1 (2015): 012312.
37) Khanna, V. K., and R. K. Nahar. "Surface conduction mechanisms and the electrical properties of Al2O3 humidity sensor." Applied surface science 28.3 (1987): 247-264.
38) Adiga, S. P., P. Zapol, and L. A. Curtiss. "Structure and morphology of hydroxylated amorphous alumina surfaces." The Journal of Physical Chemistry C 111.20 (2007): 7422-7429

39) Köck, Eva-Maria, et al. "Structural and electrochemical properties of physisorbed and chemisorbed water layers on the ceramic oxides $Y_2O_3$, YSZ, and $ZrO_2$." ACS applied materials & interfaces 8.25 (2016): 16428-16443.
40) Li, Han, et al. "High-temperature humidity sensors based on $WO_3$–$SnO_2$ composite hollow nanospheres." Journal of Materials Chemistry A 2.19 (2014): 6854-6862.